\documentclass[letter]{aa} 

\usepackage{graphicx}
\usepackage{txfonts}
\begin{document}

   \title{Shallow extra mixing in solar twins inferred from Be abundances
   \thanks{Based on observations obtained at the
   European Southern Observatory (ESO) Very Large Telescope (VLT) at Paranal Observatory, Chile (observing program 083.D-0871).}
   }


   \author{M. Tucci Maia
          \inst{1}
          \and
          J. Mel\'{e}ndez\inst{1}
          \and 
          M. Castro\inst{2}
          \and
          M. Asplund\inst{3}
          \and
          I. Ram{\'{\i}}rez\inst{4}
          \and
          T. R. Monroe\inst{1}
          \and
          J. D. do Nascimento Jr.\inst{2,5}
          \and
          D. Yong\inst{3}
          }

   \institute{Departamento de Astronomia do IAG/USP, Universidade de S\~ao Paulo, Rua do Mat\~ao 1226, 
              Cidade Universit\'aria, 05508-900 S\~ao Paulo, SP, Brazil. \email{marcelotuccimaia@usp.br}
         \and
             Departamento de F{\'{\i}}sica Te\'orica e Experimental, Universidade Federal do Rio Grande do Norte, 59072-970 Natal, RN, Brazil
         \and
             Research School of Astronomy and Astrophysics, The Australian National University, Cotter Road, Weston, ACT 2611, Australia    
         \and
             McDonald Observatory and Department of Astronomy, University of Texas at Austin, USA
         \and
             Harvard-Smithsonian Center for Astrophysics, Cambridge, Massachusetts 02138, USA}

   \date{Received ... 2014; accepted ... 2014}

 
  \abstract
   {Lithium and beryllium are destroyed at different temperatures in stellar interiors. As such, their relative abundances offer 
   excellent probes of the nature and extent of mixing processes within and below the convection zone.}
   {We determine Be abundances for a sample of eight solar twins for which Li abundances have previously been determined. 
   The analyzed solar twins span a very wide range of age, $0.5-8.2$\,Gyr, which enables us to study secular evolution of Li and Be depletion.}
   {We gathered high-quality UVES/VLT spectra and obtained Be abundances by spectral synthesis of the \ion{Be}{ii} 313\, nm doublet.}
   {The derived beryllium abundances exhibit no significant variation with age. The more fragile Li, however, 
   exhibits a monotonically decreasing abundance with increasing age. Therefore, relatively shallow extra mixing below the convection zone 
   is necessary to simultaneously account for the observed Li and Be behavior in the Sun and solar twins.}
   {}


   \maketitle
%

\section{Introduction}

The light elements lithium and beryllium are fragile, meaning
that they are destroyed at temperatures of about $2.5 \cdot 10^6$\,K and 
$3.5 \cdot 10^6$\,K, respectively (through $\alpha$ and proton captures). 
The observed abundances of Li and Be thus provide constraints for the transport of material in stellar interiors. 
To decrease the Li and Be abundances, the material has to be transported to deeper and hotter regions within the star before returning to the surface. 
In the Sun, lithium destruction 
requires temperatures somewhat hotter than those achieved at the base of the convective zone 
according to standard models of stellar evolution. The fact that the observed
photospheric Li abundance is some 150 times lower than the meteoritic value thus necessitates extra mixing below 
the convection zone. The mechanism (or mechanisms) for depleting the light elements is still debated.
Possible processes leading to extra mixing include rotation \citep{pin89}, internal gravity waves \citep{cha05},
microscopic diffusion and gravitational settling (Michaud et al. 2004),
and convective overshooting \citep{xio07}. 
Because beryllium destruction requires greater temperatures than Li, its abundance serves to constrain the extent of this extra mixing. 

The solar photospheric Be abundance has been a source of contention. 
Early work suggested that Be was depleted in the solar photosphere compared to meteorites \citep{chm75}. In solar-type stars, the Be abundances can
only be estimated through the \ion{Be}{ii} doublet at 313\,nm, a spectral region difficult to analyze because of
blends and uncertain atomic data. It has long been debated whether there is a substantial amount of missing UV opacity (e.g., Magain 1987; Kurucz 1992; Allende Prieto \& Lambert 2000).
 \cite{bal98} attempted to empirically calibrate the amount of missing continuous UV opacity by 
 enforcing the same O abundances from the OH A-X lines near the Be doublet and the OH vibration-rotation lines in the infrared.
 They inferred 
 a substantial amount of this missing opacity and also that the solar Be abundance is the 
 same as the meteoritic value within errors, a conclusion which \cite{asp04} also reached using a more sophisticated
 3D hydrodynamical  model atmosphere.
 Without properly accounting for this additional continuous opacity, the 
 Be abundance becomes underestimated,
 which leads to erroneous conclusions whether there even is a solar Be depletion; in fairness, we note, however, that \citet{chm75} concluded that
 the uncertainty in their Be abundance ($1.15 \pm 0.20$ dex) was too large to advocate any substantial Be depletion. 
 These suspicions about substantial missing UV continuous opacity were subsequently confirmed by Bell et al. (2001)
using new calculations by the Iron Project for the bound-free opacity of \ion{Fe}{i}.
More recently, \cite{tak11} studied a sample of 118 solar analogs and suggested that Be depletion 
in the Sun could be significant.

There have been several observational studies of Be in solar-type stars \citep{san04,boe09,ran10,tak11}, but none 
focused on solar twins, except for the qualitative work by \cite{tak09} on three solar twins.
The importance of solar twins is that as they have nearly solar mass and 
composition \citep{mel14b}, their evolution is similar to that of the Sun. Hence,
solar twins at different evolutionary stages in the main sequence can be used as
proxies of the Sun at different ages.
In this work, we obtain Be abundances for 
solar twins in a broad age range 
to provide constraints on Be depletion during the main sequence and thus on the extent of any extra mixing
below the convection zone.



\section{Observations and data reduction}

Spectra of eight solar twins and the Sun were obtained with the UVES spectrograph
on the 8.2m UT2 Very Large Telescope at ESO Paranal, on 29-30 August 2009.
We used the dichroic mode, obtaining simultaneous UV and optical coverage in two setups: 
{\it i)} with standard settings of 346\,nm $+$ 580\,nm; 
{\it ii)} with the standard 346\,nm and a nonestandard setting centered at 830\,nm. 
We achieved a high S/N in 
the UV because the 346\,nm setting ($306-387$\,nm) was covered in both setups. The 580\,nm standard setting
covered the optical ($480 - 682$\,nm) region, and our 830\,nm setting included the red region ($642 - 1020$\,nm). 
The UV setup was used to obtain the Be abundance, the optical setup
was used to obtain the stellar parameters and Li abundances.

We achieved a resolving power  $R = 65,000$ at $306 - 387$\,nm and $R = 110,000$ for $480 - 1020$\,nm.
The solar spectrum was also obtained through the observation of the asteroid Juno with the same spectrograph setup and served as the solar reference in our differential analysis.
The spectral orders were extracted and wavelength calibrated using IRAF, with additional data processing performed with IDL.
The reduced spectra have $S/N$ $\sim$ 100 and 1000 near the Be and Li lines, respectively. 

There are no spectral regions free from lines in the near UV spectra of solar twins. 
Hence, we employed a dedicated continuum normalization technique, which is described in Ram\'irez et al.\ (2008). 
We took advantage of the superb continuum normalization of the solar spectrum reported by Kurucz et al. (1984), 
which is used as a reference. Each order of the UVES spectra was divided by its corresponding piece from the spectrum of 
Kurucz et al.\ 
 after matching their spectral resolution, 
correcting for radial velocity offsets, and rebinning to a common wavelength sampling. 
In principle, the result should have been a smooth function corresponding to the shape of the continuum 
(the upper “envelope”) of the UVES spectra. However, because
of the finite S/N values and instrumental 
differences or small defects, this envelope had to be smoothed out using a 
100-pixel wide ($\sim$0.16\,nm) median filter. 
The UVES data were then divided by this envelope in each order. In essence, this procedure makes the UVES spectra inherit the 
continuum normalization of the spectrum of Kurucz et~al.

\section{Analysis of the Be abundances}
We used the doublet resonance lines of \ion{Be}{ii} at 313.0420\,nm  and 313.1065\,nm 
to determine Be abundances; these are the only Be lines available for observation from the ground in solar-type stars.
As the lines are in the UV region and blended by different species of atoms and molecules, the abundance determination
was carried out by means of spectral synthesis. 

\begin{figure}
\centering
\includegraphics[width=1.0\columnwidth]{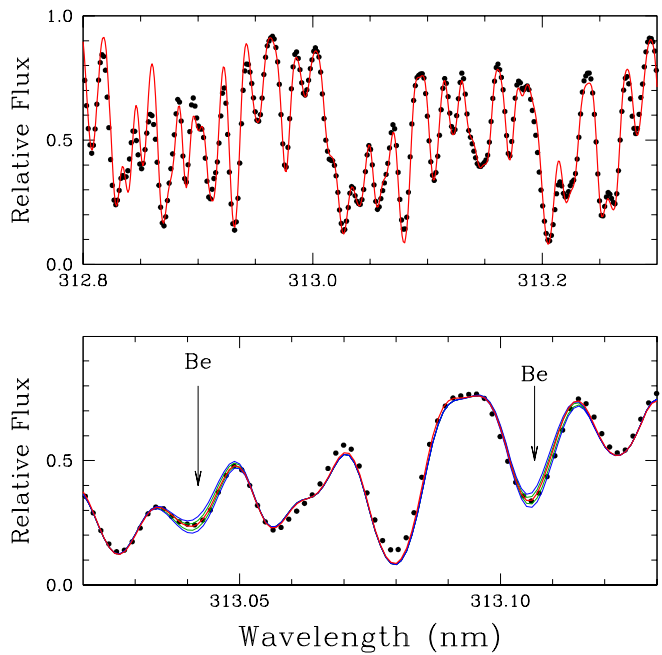}
\caption{Comparison between the observed (points) and synthetic (lines) spectra of the solar twin HIP 102152 
around the \ion{Be}{ii} lines.
The best fit is the central line, the other lines show changes in the Be abundance by $\pm$0.05 and $\pm$0.10 dex.}
\label{fig1}
\end{figure}

Our initial line list was based on the list of \cite{ash05} and
was checked with the list by \cite{pri97}, which is the list used by
\cite{tak11}. In some cases we modified the log {\it gf}-values to achieve a good agreement with the spectrum of the
Sun and of a solar analog that is severely depleted in Be (Schirbel et al. 2015, in prep.).
As the Be-poor solar analog does not show any Be lines, it was of great help to better constrain the 
lines blending the Be features in the Sun.
We emphasize that both the Sun and the Be-depleted solar analog were observed with the same
UVES setup as the solar twins and were reduced in the same way, allowing thus a reliable differential abundance analysis 
between the Sun and the solar twins.\footnote{
Our modified log {\it gf}-values of the \ion{Be}{ii} 
lines are 0.6 dex lower than those recommended by \cite{fuh10}, but we emphasize that our values 
are only valid for our internal differential analysis. In other words, they are only
valid when using our set of reduced UVES spectra,
the 2014 version of MOOG, the set of solar abundances of \cite{asp09},
and the adopted blends for the \ion{Be}{ii} region. Moreover, our Be abundances
are differential relative to the solar Be abundance of \cite{asp09}.
}
In Fig. \ref{fig1}, we show the spectral synthesis for the solar twin HIP102152. 
The weaker \ion{Be}{ii} 313.1\,nm line is less blended and offers more reliable abundances. 
Since the stronger 313.0\,nm line is heavily blended by CH and OH lines, abundances 
were determined for comparison, but not used in the final determination.

For the spectral synthesis we used the {\it synth} driver of the February 2014 version of the 1D LTE code MOOG \citep{sne73}, 
which includes continuum scattering. 
We adopted A(Be) = 1.38 dex as the standard solar Be abundance \citep{asp09}.
The model atmospheres were interpolated from the ATLAS9 Kurucz's grid
\citep{cas04} with the $T_{\rm eff}$, $\log g$, [Fe/H] and microturbulence determined 
by \citet{tal13, mel14a}, and Monroe et al. (2015; in preparation). 
\cite{asp05} and \cite{tak09}
concluded that the \ion{Be}{ii} lines are insensitive to non-LTE effects in the Sun.
Considering the similarity in stellar parameters among the solar twins and the differential nature of the analysis, 
any differential NLTE corrections would probably be vanishingly small \citep[e.g.,][]{mel12}, hence no corrections were applied.

To determine the macroturbulence line broadening, we first analyzed the line profiles of 
the \ion{Fe}{i} 602.7050 nm, 609.3644 nm, 615.1618 nm, 616.5360 nm, 670.5102 nm and \ion{Ni}{i} 676.7772 nm lines in the Sun;  
the syntheses also included a rotational broadening of $v \sin i = 1.9$\,km s$^{\rm -1}$ \citep{bru84,saa97},
and the instrumental broadening.
The macroturbulent velocity found for the Sun is $V_{{\rm macro}, {\sun}} = 3.6$\,km s$^{\rm -1}$.
For the solar twins, we estimate the macroturbulence following \cite{mel12} (average of equations E.2 and E.3):
$V_{\rm macro, star} = V_{\rm macro, \sun} + (T_{\rm eff} - 5777)/486.$

With the macroturbulence fixed, $v \sin i$ was estimated for the solar twins sample 
by fitting the profiles of the six lines mentioned above, also including the instrumental broadening. 
Finally, the best-fitting Be abundances were estimated 
using a $\chi^2$-procedure. The estimated macroturbulence, $v \sin i$ and Be abundances for the entire sample can be found in Table 1.

We estimated the errors considering both observational and systematic uncertainties. 
The observational errors are due to uncertainties of the continuum placement and $S/N$
(the synthetic spectra were shifted vertically within the allowed noise of the 
observed spectrum; the abundance variation due to this shift was adopted as the observational error). 
For the systematic errors we considered the errors in the  stellar parameters. 
Both observational and systematic errors were added in quadrature.

Thanks to the high internal precision of the atmospheric parameters derived in our solar twin stars,
we were able to employ standard isochrone techniques to measure reliable relative ages for these objects.
We adopted the ages previously derived by our group (Monroe et al. 2013, 2015; Melendez et al. 2014a) using
the algorithm described in Ram\'{\i}rez et~al.\ (2013, 2014), which computes the age probability distribution
function by comparing the location of the star on the $T_\mathrm{eff}, \log g, \mathrm{[Fe/H]}$ parameter
space with the values predicted by theory. For the two youngest stars, we also used
other age indicators, as described in Monroe et al. (2015). 
In short, we adopted a rotational age (0.5$\pm$0.2 Gyr) for HD 20630, which excellently
agrees with 
\cite{rib10}, who obtained 0.6$\pm$0.2 Gyr
using different indicators.
For HD 202826 a rotation period
is not available to estimate its age, hence we used its chromospheric activity, X-ray luminosity, and isochrones.
From the isochrones we derive an average mass 
for the sample stars of 1.01 $\pm$ 0.03 M$_\odot$, that is, solar within the errors, reinforcing thus the
use of solar twins as proxies of the Sun at different ages. 
The estimated ages are given in Table 1. \footnote{Note that our ages are differential 
and that our errors are only internal. 
We caution that the ages may turn out to be slightly older or younger depending 
on the choice of isochrones, making the age range covered by the sample slightly wider or narrower, but the relative
ages are reasonably well constrained,
as shown for example in Fig. 5 of \cite{mel14a}, where the relative ages between 
the Sun and the solar twin 18 Sco are consistent for two different sets of isochrones.}

\section{Discussion}

\begin{figure}
\centering
\includegraphics[width=1.0\columnwidth]{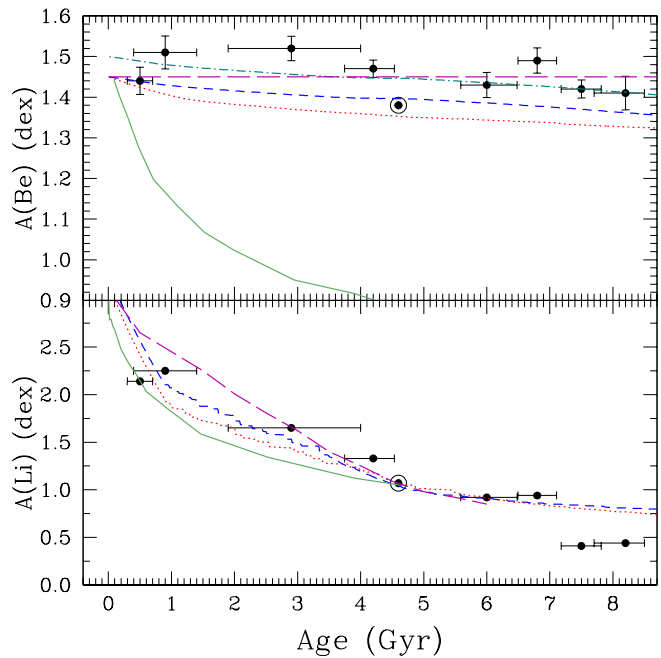}
\caption{Be (top) and Li (bottom) abundances vs. age.
For the models of Be depletion, we adopt an initial meteoritic A(Be) = 1.45 dex, which is based 
on the value by \cite{lod03} (1.41 dex), plus the 0.04 dex gravitational settling effect over 4.5 Gyr.
The models of Li depletion were normalized to the solar Li abundance.
The green solid lines are the models by \cite{pin89}, the red dotted lines represent the models by \cite{nas09}, 
the blue dashed lines are the modified models of \cite{nas09} (see text), and 
the purple long-dashed lines are predictions of Be depletion by  \cite{xio07} and Li depletion by \cite{xio09}. 
The teal dotted-dashed line (top panel)
is the same Be depletion model of \cite{nas09} shown by the red dotted lines, but with a higher initial 
A(Be)= 1.50 dex (rather than A(Be) = 1.45 dex). The additional
0.05 dex is to compensate for the refractory depletion of Be in the Sun (see text).}
\label{fig2}
\end{figure}


Figure \ref{fig2} (top panel) shows the measured Be abundances in our solar twins as a function of stellar age.
The scatter in the Be content is very small: $0.04$\,dex.  This is significantly less than in previous studies;
for example,  \cite{tak11} analyzed Be in a large sample of solar analogs and found a dispersion of 0.2 dex. 
A linear fit to our data using the error bars in both age and Be abundances gives a slope of $-8.09 \cdot 10^{-3} \pm 4.17 \cdot 10^{-3}$\,dex/Gyr, which is a shallow
trend at best. 
For comparison, we show in the bottom panel the corresponding NLTE Li abundances for the same sample stars
(Monroe et al. 2015). A fit to the Li data gives a slope of $-0.23 \pm 0.01$\,dex/Gyr.
Clearly, there is a steep Li depletion, but the mixing processes responsible for the destruction of lithium 
must be relatively shallow and cannot transport material to deeper regions where significant Be destruction can occur.

In standard evolution models of the Sun, the depletion of the light elements Be and Li is expected to 
occur only below the convective zone and thus the surface abundances should remain unchanged, obviously
in contrast to the observational evidence in the case of Li. 
Only for less massive stars do standard stellar models predict significant Be depletion as a result of the deeper 
convective zone \citep{san04}. 
Owing to the monotonically increasing Li depletion with age  \citep{tal13}, most Li destruction occurs
on secular timescales during the main sequence.
Since standard models cannot predict this behavior, there must be extra mixing below the convection zone
that brings material down to sufficiently large depths and temperatures, and then back into the convection zone. 
In Fig. \ref{fig2}, four different predictions 
for Be and Li depletion are compared. The model by \cite{pin89} considers rotationally 
induced mixing and depletes Li reasonably well, but 
depletes far too much Be (green solid line).
The model of  \cite{nas09} includes extra mixing due to diffusion (including gravitational settling) and rotation. 
It reproduces Li in solar twins well, but Be depletion is somewhat larger than observed
(red dotted lines). We have modified this model to calibrate the amount of meridional circulation,
achieving a steep and a shallow Li and Be depletion, respectively (blue dashed lines).
The new model has meridional circulation with a lower efficiency, hence destroying less Be and Li,
and we increased the turbulent diffusion coefficient of the tachocline 
below the convective zone to destroy more Li (without affecting Be). 
Finally, the models of \cite{xio07,xio09} incorporate convective overshoot as well as gravitational settling. 
They reproduce the Li depletion with age but do not deplete Be (purple long-dashed line), 
unlike the observations, which seems to suggest a shallow depletion of Be.

The initial (zero-age) Be abundance for the model predictions assumes them
to be equal to the meteoritic values. For the meteoritic abundance, 
\citet{lod03} recommended A(Be)$=1.41 \pm 0.08$ dex, which was subsequently revised to A(Be)$=1.32 \pm 0.03$ dex by \citet{lod09}. 
These are indirect measurements, however, because Be is difficult to measure in carbonaceous chondrites of type CI,
which are the least modified meteorites and thus the preferred choice when inferring the primordial solar system abundances.
Instead, the meteoritic value was estimated from the relative abundances of refractory elements in CM and CV chondrites in which
Be has been measured in a couple of cases. The relatively large uncertainty for Be for being meteorites reflects
this indirect procedure. 

Since the absolute abundance scales for meteorites are set by enforcing that 
the photospheric and meteoritic Si abundances are equal \citep{asp00, asp09}
and all elements heavier than hydrogen have experienced gravitational settling in the Sun 
over the past 4.5\,Gyr \citep[see discussion in ][]{asp09}, a more appropriate initial Be value for these model
predictions would thus be 0.04\,dex higher than the abundances recommended by \citet{lod03} and \citet{lod09}. 
Our solar twins data would seem to suggest that the higher meteoritic Be abundance (1.41 dex)
is more appropriate. We thus added 0.04 dex to this value, adopting A(Be) = 1.45 dex as the
initial Be abundance in Fig. \ref{fig2}. 
The Sun may be slightly less abundant in Be by $\sim0.05$\,dex for its age compared with
other similar solar twins (Fig. \ref{fig2}), but a larger number of solar twins would be
required to confirm this impression. 
Perhaps the  somewhat lower solar Be abundance could arise simply 
because the Sun is poor in refractories \citep{mel09};
Be has a condensation temperature of $T_{\rm cond}$ = 1452 K \citep{lod03}, which means that it is a refractory element. 
The Sun is probably deficient in refractories as a result of the formation of rocky planets in the solar system \citep{mel09}.
For its $T_{\rm cond}$, we estimate that Be should be depleted in $\sim$0.05 dex in the Sun.
Interestingly, if we consider an initial A(Be) = 1.5 dex owing to the refractory depletion of Be, the modified model
by \cite{nas09} reproduces the Be abundances of most solar twins
well (dotted-dashed line in Fig. \ref{fig2}).

Recently, \cite{adi14} suggested that the depletion of refractory elements in the Sun
relative to solar twins could be an age effect. If this interpretation is
correct, and because Be is a refractory element, we should have found that beryllium in 
solar twins older than the Sun (4.6 Gyr) is depleted relative to the Sun, because 
\cite{adi14} analyzed their solar analogs relative to the Sun. 
However, this is not what we observe in Fig. \ref{fig2}. Thus, our results seem to
be in conflict with the interpretation by \cite{adi14}.


\section{Conclusions}

We presented the first detailed study of beryllium abundances in solar twins covering a broad range of ages (0.5 - 8.2 Gyr). 
Our analysis revealed that the Be abundance is relatively constant with age, with a scatter of only 0.04 dex and a weak, if any, trend with age. This is
in contrast to the large observed depletion of Li with age \citep{bau10,tal13,mel14b}, showing that the transport mechanisms are deep enough
to reach the region where Li is burned, but not deep enough to reach the higher temperatures needed to burn Be.
Our Li and Be results provide stringent constraints on  stellar models and nonstandard mixing processes beyond treating convection through 
the mixing length theory.

\begin{acknowledgements}
We thank Johanna F. Jarvis for sharing her Be line list.
MTM thanks for support by CNPq (142437/2014-0). JM thanks for
support by FAPESP (2012/24392-2).
MA and DY acknowledge financial support from the Australian Research Council (grant DP120100991). 
\end{acknowledgements}

\begin{table}
\caption{Be abundances and errors for the eight solar twins and the Sun, together with the inferred macroturbulence, $v \sin i,$ and ages of the stars.}
\label{parameters1}
{\centering
\begin{tabular}{lcccc} 
\hline\hline                
 {Star}& $V_{\rm macro}$/$v \sin i$ & A(Be) & param$^{\rm a}$/obs$^{\rm b}$/total$^{\rm c}$& Age\\
 {} & (km s$^{\rm -1}$)  &  (dex) & (dex) & (Gyr)\\
\hline 
HD20630   & 3.5/4.2 &1.44& 0.03/0.03/0.04&$0.5_{\rm -0.2}^{\rm +0.2}$\\
HD202628  & 3.7/2.4 &1.51& 0.01/0.04/0.04&$0.9_{\rm -0.5}^{\rm +0.5}$\\
HIP30502  & 3.5/1.6 &1.43& 0.01/0.03/0.03&$6.0_{\rm -0.4}^{\rm +0.5}$\\
HIP73815  & 3.6/1.7 &1.49& 0.01/0.03/0.03&$6.8_{\rm -0.3}^{\rm +0.3}$\\
HIP77883  & 3.4/1.8 &1.42& 0.01/0.02/0.02&$7.5_{\rm -0.3}^{\rm +0.3}$\\
HIP89650  & 3.8/1.7 &1.47& 0.01/0.02/0.02&$4.2_{\rm -0.5}^{\rm +0.3}$\\
18Sco     & 3.7/2.0 &1.52& 0.01/0.03/0.03&$2.9_{\rm -1.0}^{\rm +1.1}$\\
HIP102152 & 3.5/1.8 &1.41& 0.01/0.04/0.04&$8.2_{\rm -0.5}^{\rm +0.3}$\\
Sun       & 3.6/1.9 &1.38& 0.00/0.01/0.01& 4.6\\
\hline                                 
\end{tabular}
}
\\
$^{\rm a}$ errors due to stellar parameters \\
$^{\rm b}$ observational errors \\
$^{\rm c}$ quadric sum of the observational and systematic errors
\end{table}


\end{document}